\documentclass[a4paper,11pt]{article}

\usepackage[T1]{fontenc}
\usepackage[utf8]{inputenc}
\usepackage{lmodern}

\usepackage{amsmath,amssymb,amsfonts}
\usepackage{graphicx}
\usepackage{cite}
\usepackage{hyperref}
\usepackage[a4paper,margin=3cm]{geometry}

\newcommand{\orh}{}
\newcommand{\mb}{}

\title{Machine Learning Surrogate Modeling for Homogenization of Hyperelastic Materials with Boolean Microstructures}

\author{
Matthias Brändel\thanks{
Technische Universität Bergakademie Freiberg, Institut für Numerische Mathematik und Optimierung,
Fakultät für Mathematik und Informatik, 09596 Freiberg, Germany.
}
\and
Oliver Rheinbach\thanks{
Technische Universität Bergakademie Freiberg, Institut für Numerische Mathematik und Optimierung,
Fakultät für Mathematik und Informatik, 09596 Freiberg, Germany.
Corresponding author: \texttt{oliver.rheinbach@math.tu-freiberg.de},
phone +49\,3731\,393279.
}
}

\date{}

\begin{document}

\maketitle

\begin{center}
\small
This is an author-prepared preprint corresponding to the manuscript submitted to PAMM. The formatting has been adapted for arXiv; the scientific content, results, and conclusions are unchanged.
\end{center}

\begin{abstract}
Data-driven surrogate models offer an efficient alternative to repeated numerical homogenization of heterogeneous materials.
In this contribution, a supervised learning approach is presented for predicting effective Lam\'e parameters of hyperelastic composites from low-dimensional microstructural descriptors.
The data set is based on previously published numerical homogenization results for ensembles of two-phase stochastic microstructures generated by planar Boolean models, covering variations of inclusion shape, phase contrast, and area fraction, see Brändel, Brands, Maike, Rheinbach, Schröder, Schwarz and Stoyan (2022).
A neural network is trained on combinations of scalar and curve-valued statistical descriptors, including the area fraction, a derived scalar shape descriptor $\tau$, the two-point correlation function $S_2(r)$, and the lineal-path function $\ell(z)$.
Additional data representing limiting cases of the parameter space are incorporated to stabilize training and improve extrapolation behavior.
The surrogate is evaluated by leave-one-grain-type-out cross-validation in order to assess generalization to unseen grain geometries.
Numerical results demonstrate that additional descriptors can reduce relative errors.
A predictor trained with $\tau$ and $S_2(r)$ provides a compact representation with good quantitative accuracy and regular dense response behavior.
Adding the lineal-path function $\ell(z)$ further reduces the error at the available data points, indicating that it is a promising additional descriptor; however, dense post-training response evaluations show that improved pointwise accuracy does not automatically guarantee physically admissible behavior between sampled parameter values.
This motivates future work on physically constrained surrogate models, loss formulations, bounded output parametrizations, and a more systematic representation of curve-valued geometric descriptors.
\end{abstract}

\section{Introduction}
The macroscopic response of heterogeneous materials is governed not only by the properties of the individual phases, but also by their spatial arrangement at the microscale.
Computational homogenization provides a general numerical framework for this scale transition, but for random microstructures it becomes costly: reliable effective quantities require repeated simulations on large samples or ensembles of microscopic boundary value problems.

Random microstructures can be modeled systematically using models from stochastic geometry.
Such models are attractive for parametric studies because they provide direct access to statistical descriptors of the microstructure.
These descriptors are closely related to classical approaches for characterizing random heterogeneous materials and their effective properties.

Recent machine-learning approaches in computational homogenization include surrogate localization models for FE$^2$ (Finite Element squared) computations, e.g., \cite{2026Klawonn}, direct prediction of microscopic fields, and convolutional neural networks for effective properties from voxelized microstructures \cite{2020Hatano,2021Mianroodi,2023Eidel}.
Our approach to use known descriptors directly as input for a neural network is complementary to these image-based surrogate models.

The microstructure generation, finite-strain homogenization procedure, and statistical assessment of the representative volume element are not the main contribution of this work, but provide the data basis for the surrogate model. They are therefore summarized only to the extent required to make the construction of the training data transparent; further details are given in \cite{2022Braendel,2023Braendel}.

\section{Microstructure}

The microstructures are modeled as two-phase structures on the planar domain. We use realizations of Boolean models to build our microstructure samples, where the area covered by the Boolean model is phase 1 and the remainder is phase 2.

\begin{figure}
\centering
    \includegraphics[width=0.25\linewidth]{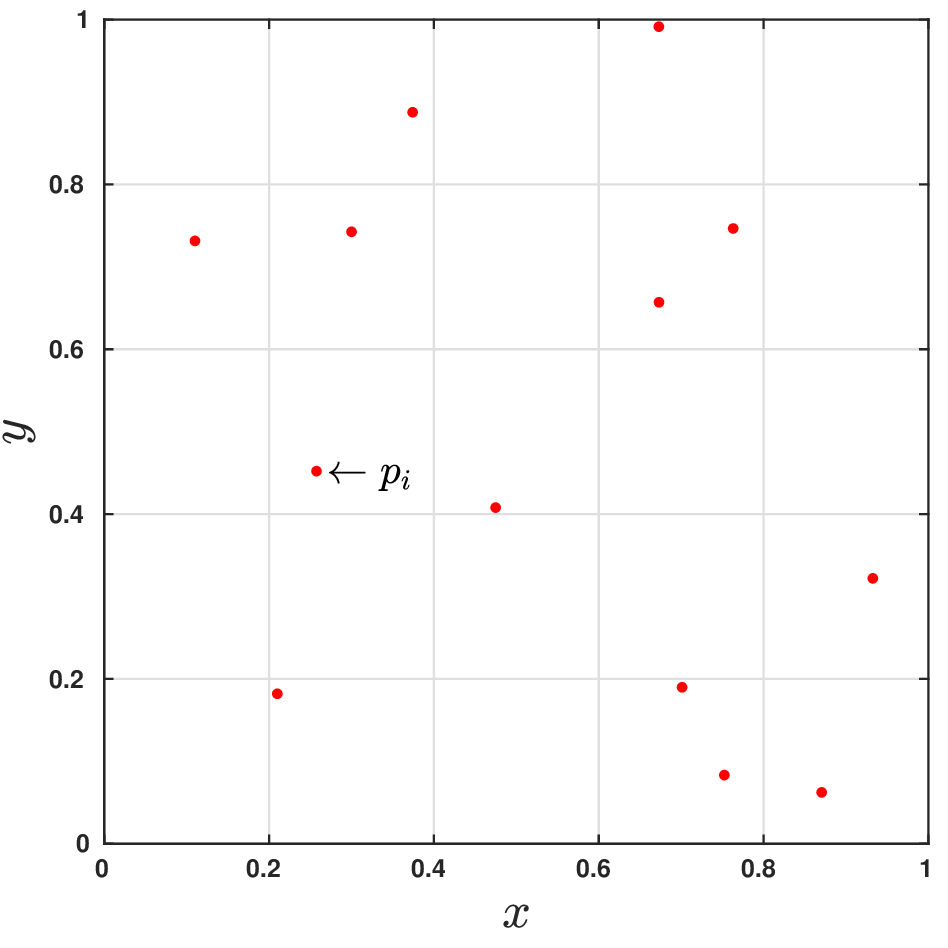}~(a)
    \includegraphics[width=0.25\linewidth]{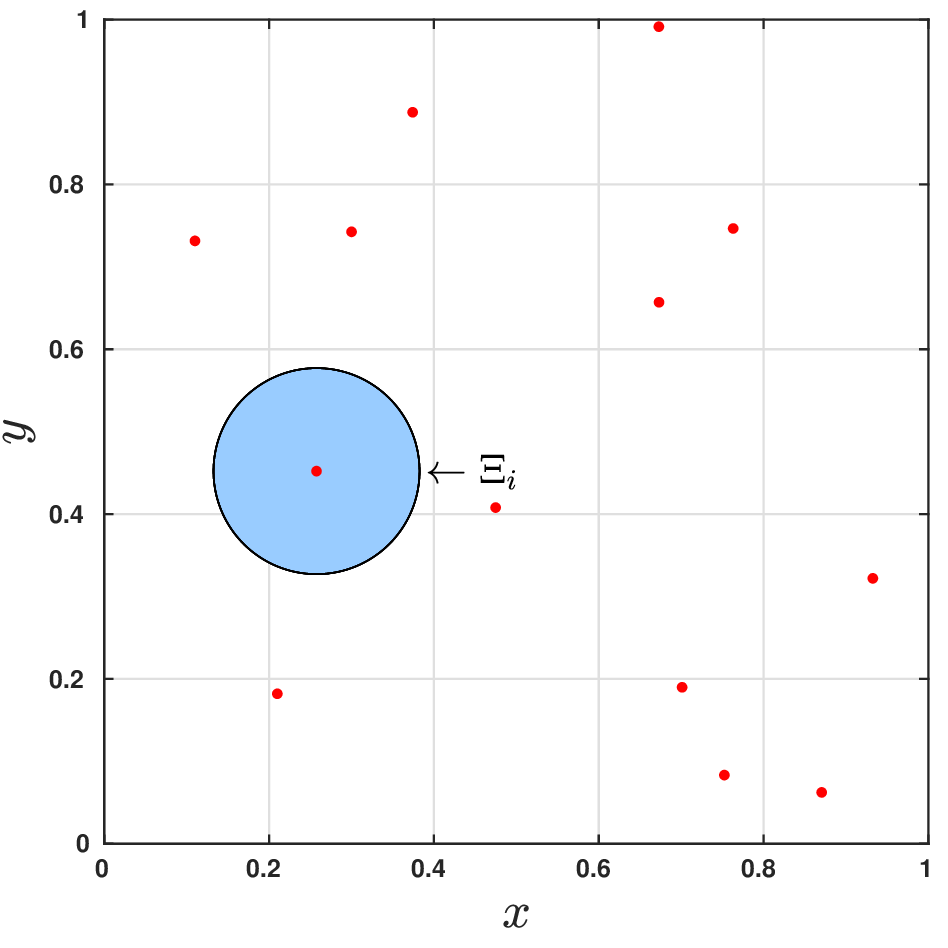}~(b)
    \includegraphics[width=0.25\linewidth]{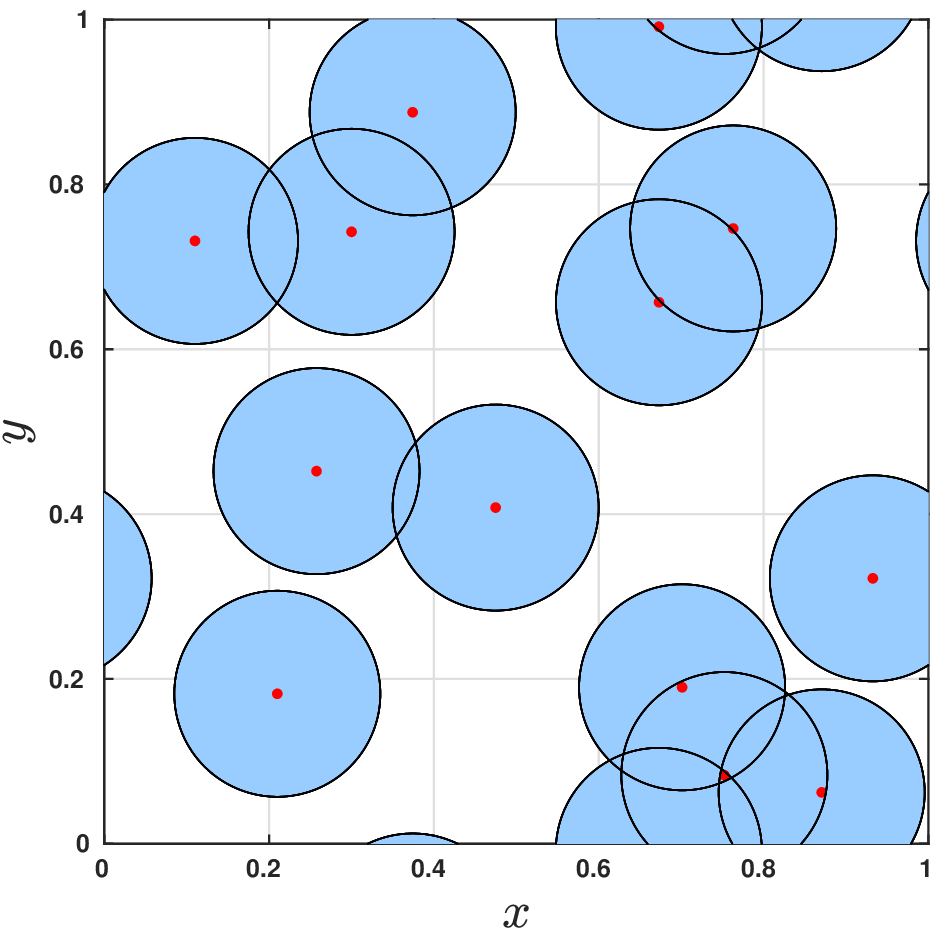}~(c)
\caption{A realization of the Poisson point: p points $p_i$ (a), placing a circular grain $\Xi_i$ (b), and a microstructure sample with overlapping grains (c).}
\label{fig:booleanmodelrealizationexample}
\end{figure}

The Boolean model is a standard germ-grain model, see, e.~g.~\cite{2013Chiu}, and has been subject to computational homogenization in linear, e.~g., \cite{2009Willot}, and nonlinear theory, e.~g., \cite{2018Willot,2022Braendel}. The germs $p_i$ form a homogeneous Poisson point process and the grains $\Xi_i$ are independent identically distributed compact sets in $\mathbf{R}^2$, compare Figure \ref{fig:booleanmodelrealizationexample} (a), (b).
In this work, this grain distribution is degenerate for each parameter combination: all grains are congruent copies of the typical grain $\Xi_0$.
The corresponding Boolean model $\Xi$ is a random set
\begin{equation}
    \Xi = \bigcup\limits_{i=1}^{\infty} \left(\Xi_i+p_i\right).
\end{equation}

The statistical properties of the model are determined by the Poisson intensity $\theta$ and the distribution of $\Xi_0$.
It is ergodic, that is, spatial averages of increasingly large windows converge to ensemble averages. A homogeneous Poisson point process is isotropic. Isotropy of the Boolean model is obtained by uniformly rotating the grains.
Note that that the grains can overlap and form clusters.

\subsection{Microstructure Generation}
The Boolean model is defined over the whole Euclidean plane, whereas numerical homogenization requires a bounded computational domain. We therefore sample the model in a window $\Omega=(0,1)^2$. A direct restriction to $\Omega$ would truncate grains intersecting the boundary. In this work, grains crossing one side of $\Omega$ are continued periodically on the opposite side, so that the generated microstructure is compatible with a periodic observation window; see Figure \ref{fig:booleanmodelrealizationexample} (c).

\begin{figure}
\begin{center}
\setlength{\tabcolsep}{2pt}
\renewcommand{\arraystretch}{0}
\begin{tabular}{ccc}
    \includegraphics[width=0.25\linewidth]{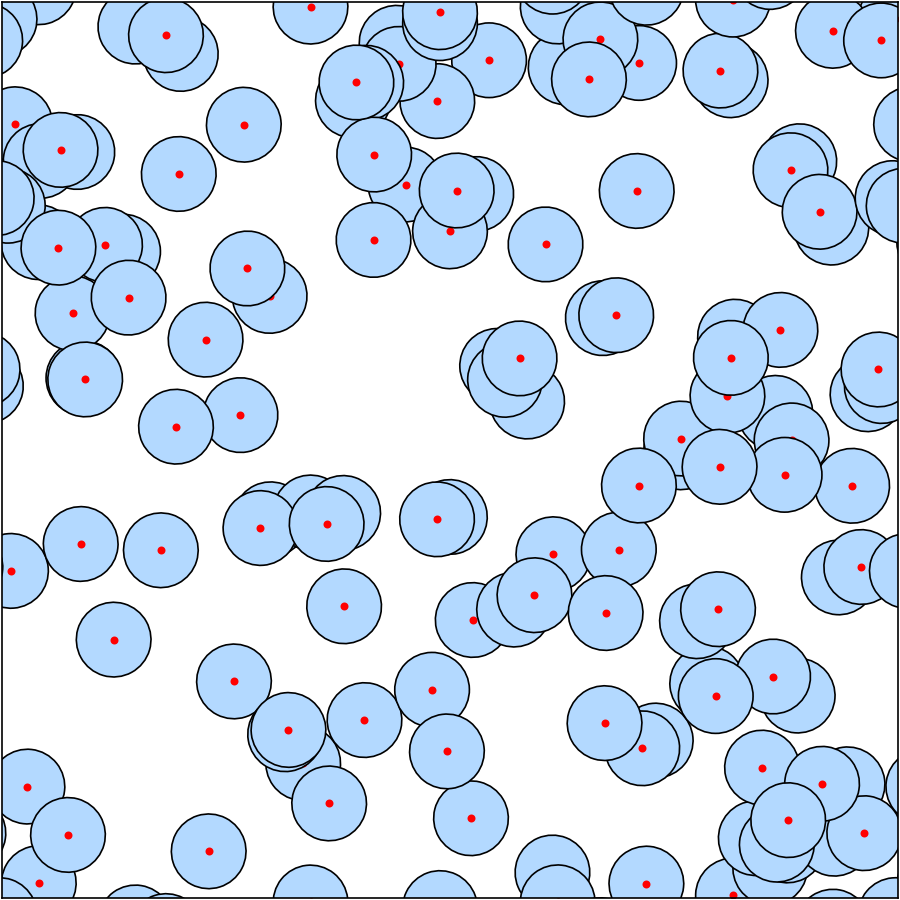}~(a) &
    \includegraphics[width=0.25\linewidth]{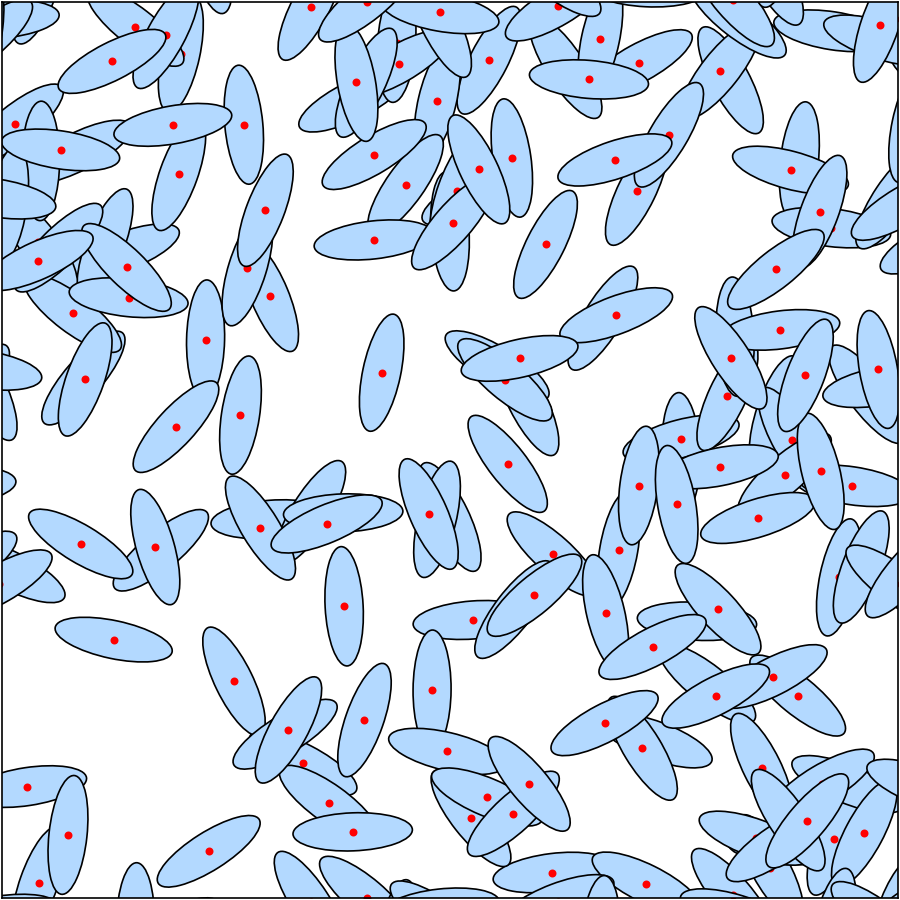}~(b) &
    \includegraphics[width=0.25\linewidth]{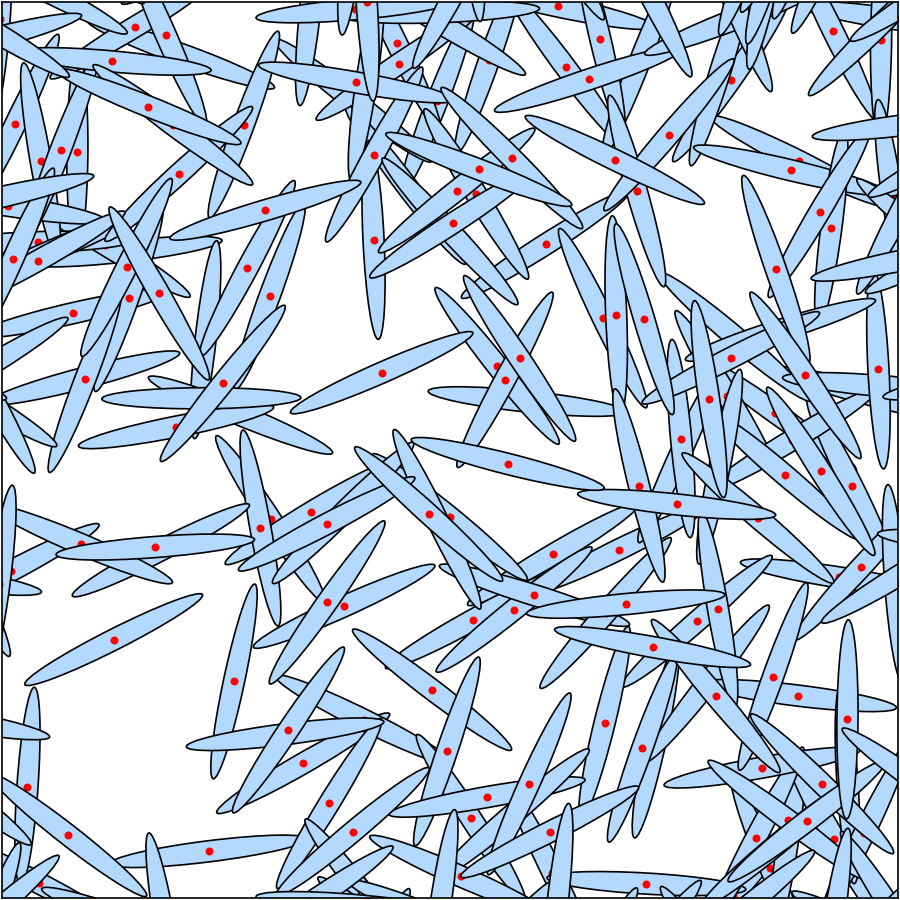}~(c) \\[5pt]
    \includegraphics[width=0.25\linewidth]{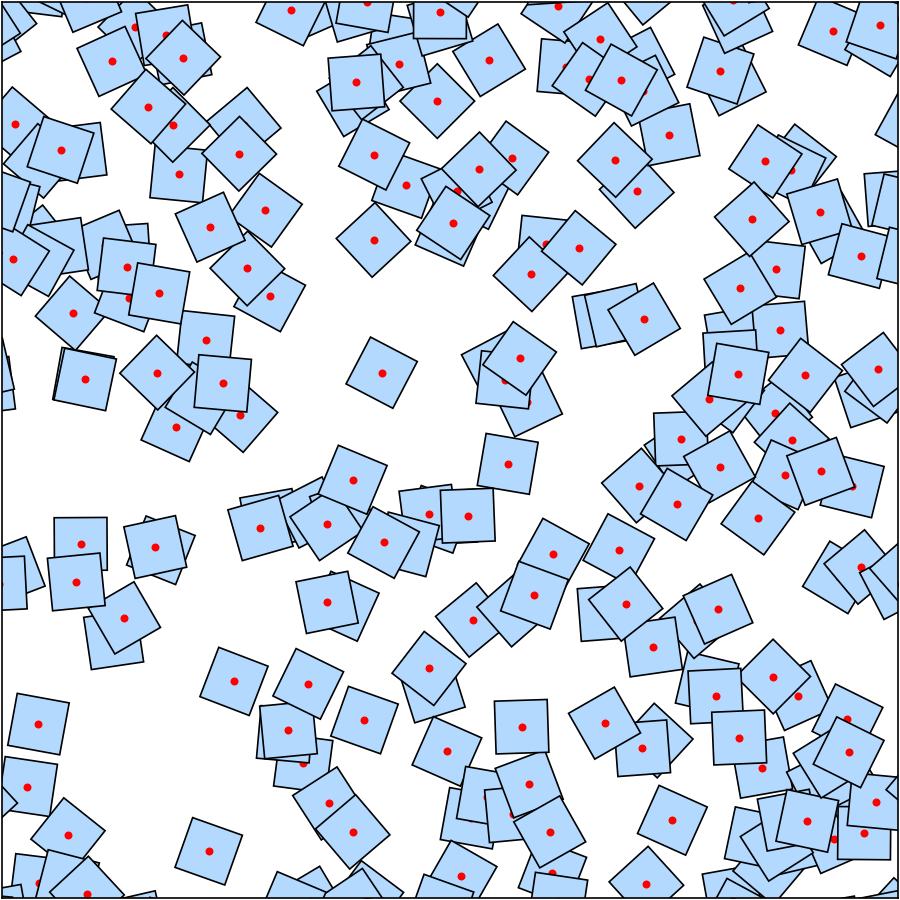}~(d) &
    \includegraphics[width=0.25\linewidth]{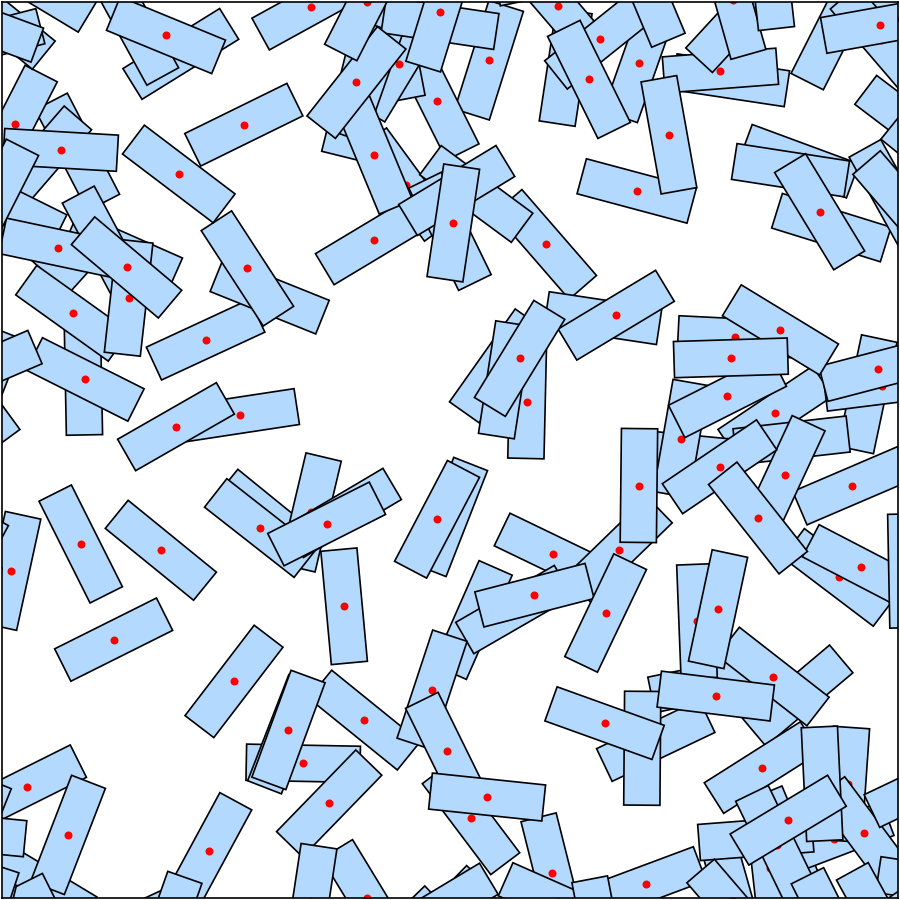}~(e) &
    \includegraphics[width=0.25\linewidth]{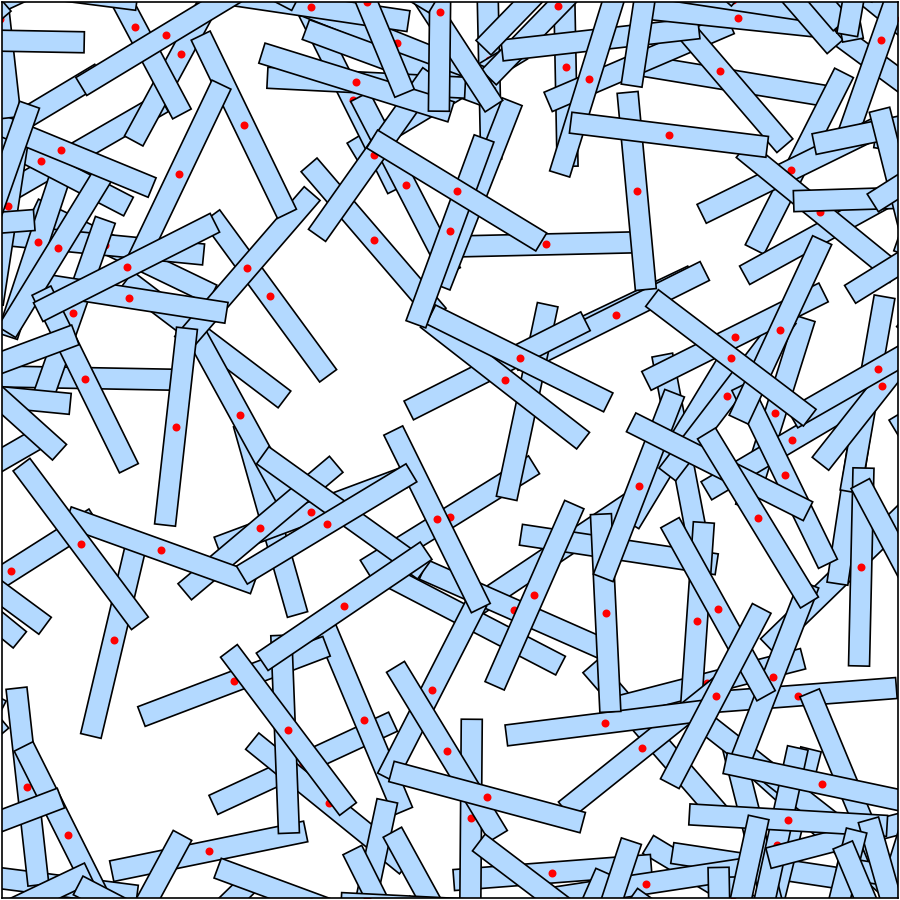}~(f) \\[5pt]
    \includegraphics[width=0.25\linewidth]{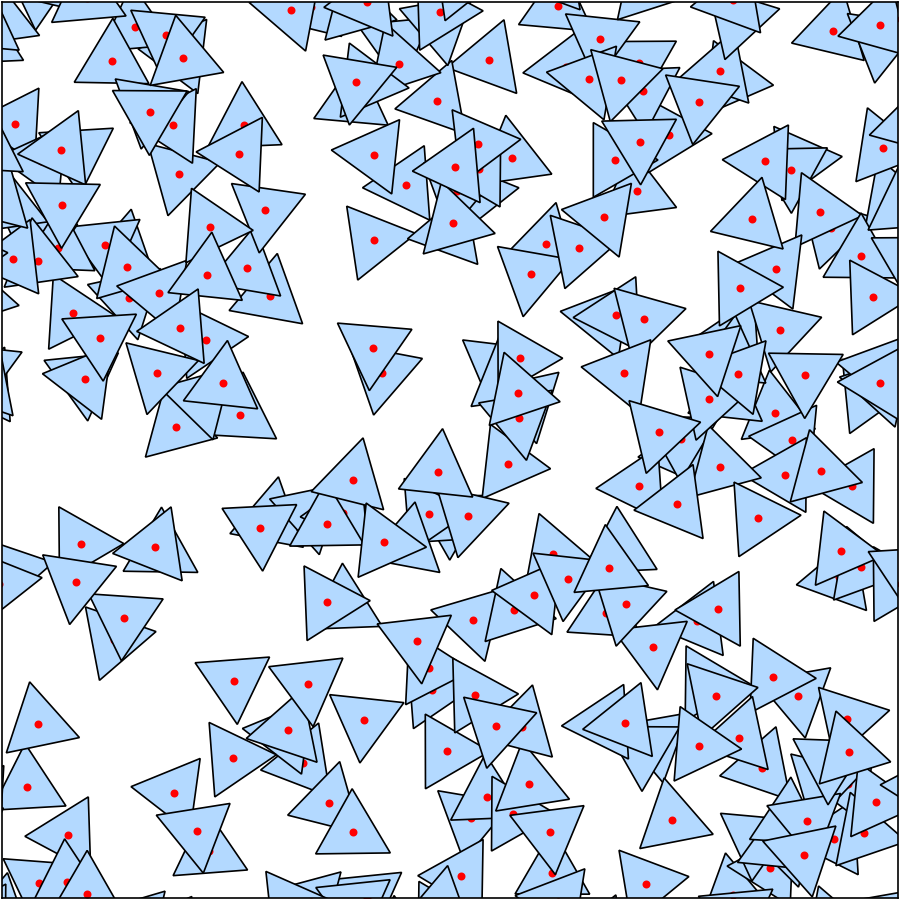}~(g) &
    \includegraphics[width=0.25\linewidth]{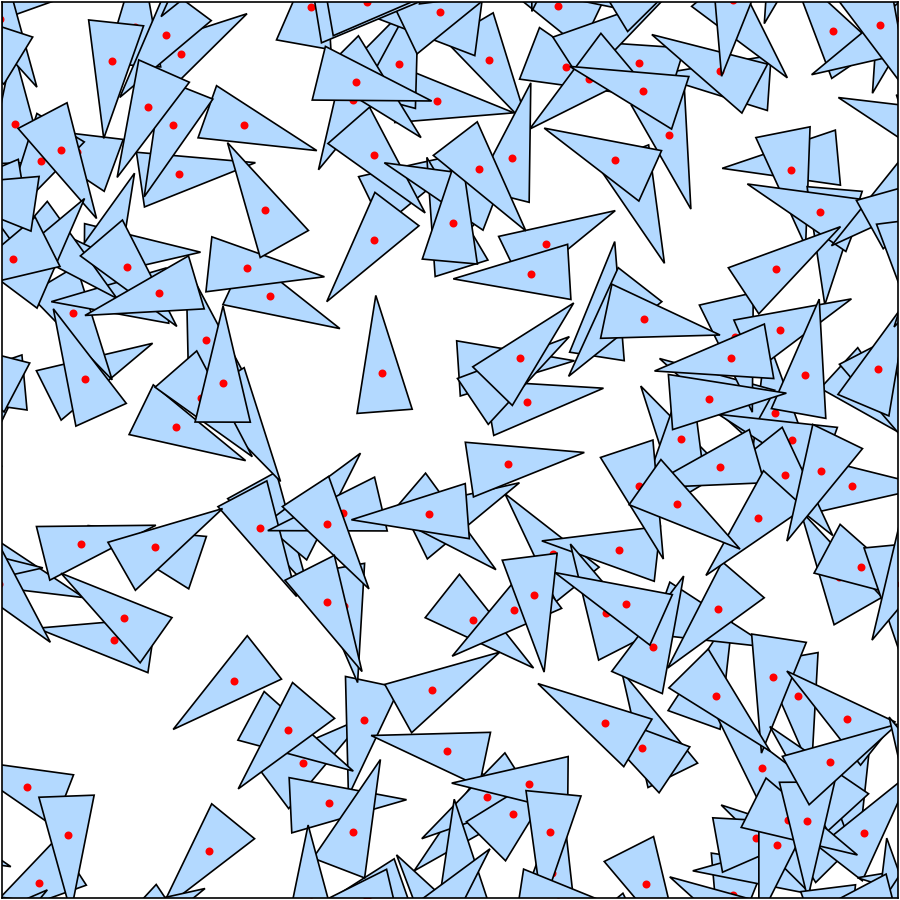}~(h) &
    \includegraphics[width=0.25\linewidth]{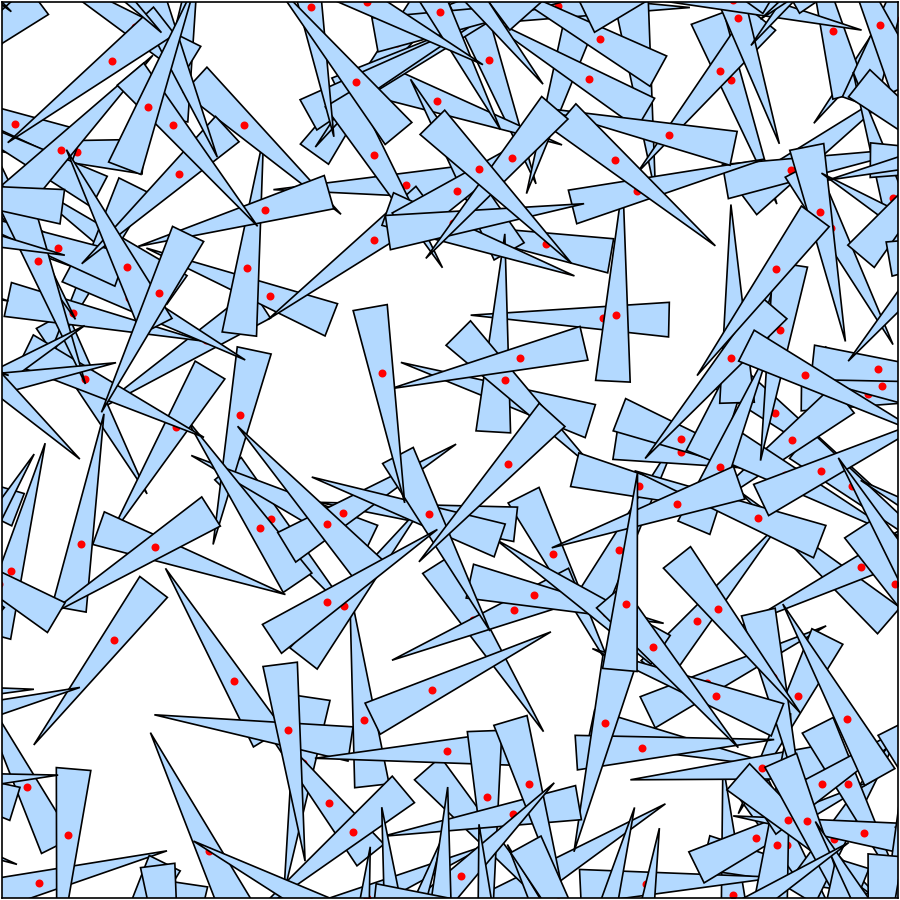}~(i)
\end{tabular}
    \caption{Boolean model microstructure samples for different grain types at area fraction $A_A=0.5$: circles $D$ (a), ellipses $E_1$ (b), elongated ellipses $E_3$ (c), quads $Q$ (d), rectangles $R_1$ (e), elongated rectangles $R_3$ (f), equilateral triangles $S$ (g), isosceles triangles $T_1$ (h), elongated triangles $T_3$ (i).}
\label{fig:grain-types}
\end{center}
\end{figure}

As grain types we use ellipses, rectangles, and triangles of aspect ratios $\{1,1:\pi,1:3\pi\}$ as well as $1:16$ only for rectangles, hence circles $D$, ellipses $E_1$, elongated ellipses $E_3$, quads $Q$, rectangles $R_1$, elongated rectangles $R_3$, very elongated rectangles $R_5$, equilateral triangles $S$, isosceles triangles $T_1$, elongated triangles $T_3$. In Figure \ref{fig:grain-types} microstructure samples with these grain types are shown.

\begin{figure}
\begin{center}
\begin{minipage}{0.85\linewidth}
\includegraphics[width=\linewidth]{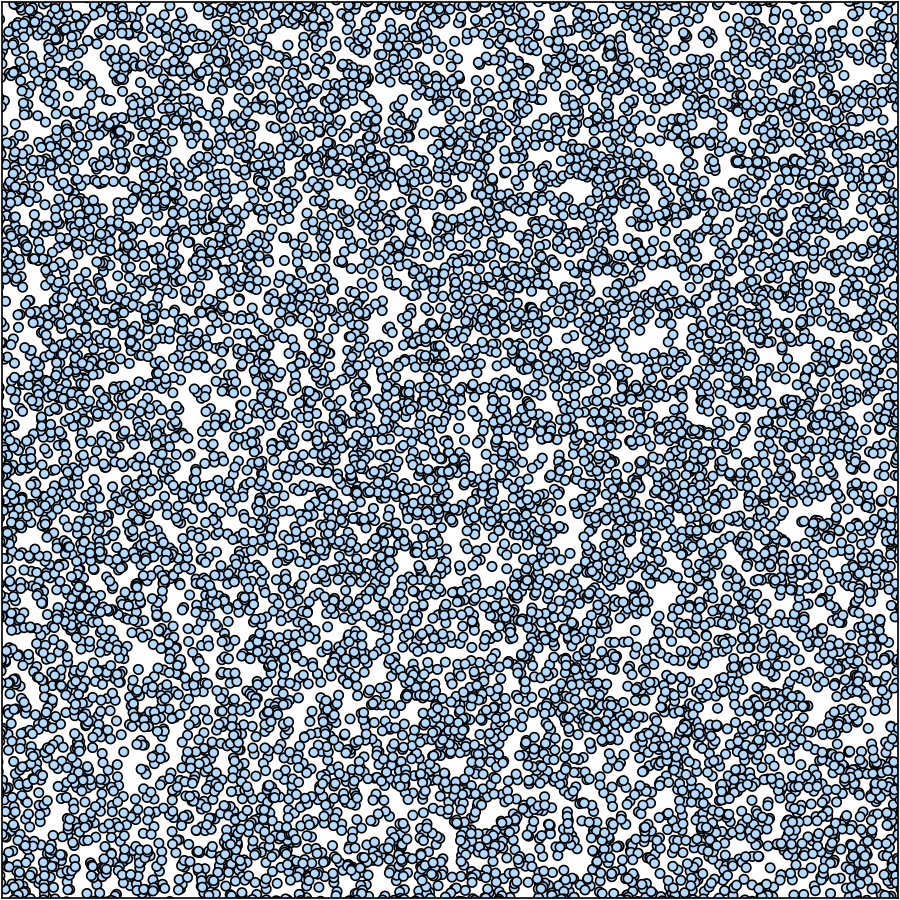}
    \caption{Boolean microstructure sample for circles as grains at a sample size of $\delta=96$ and an area fraction $A_A=0.6$.}
\label{fig:sampled96}
\end{minipage}
\end{center}
\end{figure}

We increase the size of the microstructure sample not by increasing the size of the observation window, but by reducing the size of the grains while increasing their number. For this we use the sample size $\delta$ which is the ratio of the length of the observation window $L$ and the diameter of the circle circumscribing a grain $d_{B(\Xi_0)}$,
\begin{equation}
    \delta = \frac{L}{d_{B(\Xi_0)}}.
\end{equation}
With increasing sample size the statistical error reduces.
An example for a microstructure sample from an ensemble from which we compute effective material parameters is shown in Figure \ref{fig:sampled96}.

We represent a given Boolean model by an ensemble of $n$ microstructure samples.
To estimate the size of the representative volume element, we perform numerical homogenization on sequences of ensembles for the same grain type with increasing sample sizes.
For small sample sizes, the statistical error is very high.
We therefore use many more samples in the ensemble.
To obtain a constant error in the spatial averages over the ensemble the ensemble size $n$ should be proportional to $\delta^{-2}$.
To reduce computational cost we choose $n\propto\delta^{-\sqrt{2}}$. It holds however that $n\geq200$.

\subsection{Statistical Descriptors}
%
Statistical descriptors are quantitative measures that characterize the morphology of a microstructure.
%
Using statistical descriptors as neural-network inputs offers an alternative to image-based approaches. In the latter, convolutional layers implicitly learn feature representations from microstructure images, which may not align with classical descriptors.

The breadth of available statistical descriptors is reflected in the monographs \cite{2000Ohser,2002Torquato}, which collect approaches from mathematical morphology, stereology, spatial statistics, stochastic geometry, and the theory of random heterogeneous materials.
Accordingly, descriptor-based microstructure characterization should be understood as an extensive toolbox, ranging from standard quantities such as volume fractions, correlation functions, chord-length and contact distributions, and Minkowski functionals to specialized descriptors developed for particular classes of morphologies and structure-property relations.

In this work, we additionally restrict the Boolean model to isotropically distributed convex grains. Under these assumptions, several statistical descriptors admit analytic expressions in terms of the model parameters and therefore need not be estimated from image data.

We use the \textit{area fraction} $A_A$,
\begin{equation}
    A_A = 1 - \exp\left(-\theta A\right),
\end{equation}
where $A$ is the mean area of $\Xi_0$, the \textit{specific boundary length} $L_A$,
\begin{equation}
    L_A = \theta\exp\left(-\theta A\right)L,
\end{equation}
where $L$ is the mean circumference of $\Xi_0$, the \textit{Euler characteristic density} $\chi_A$,
\begin{equation}
    \chi_A = \theta\exp\left(-\theta A\right) \left(1-\theta\frac{L^2}{4\pi}\right),
\end{equation}
the \textit{lineal-path function} $\ell(z)$,
\begin{equation}
    \ell(z) = A_A\exp\left(-\frac{L_A}{\pi A_A}z\right),
\end{equation}
the \textit{chord-length density} $p(z)$,
\begin{equation}
    p(z) = \frac{L_A}{\pi A_A}\exp\left(-\frac{L_A}{\pi A_A}z\right),
\end{equation}
and the \textit{two-point correlation function} $S_2(r)$,
\begin{equation}
    S_2(r) = 1 - 2\exp\left(-\theta A\right) + \exp\left(-\theta A_{K\cup K_r}\right),
\end{equation}
where $A_{K\cup K_r}$ is the mean union area of two grains at distance $r$, whose closed form is only available for circles. For the other grain types, we approximate the mean overlap area by angular sampling and obtain the mean union area from $A_{K \cup K_r} = 2A - A_{K \cap K_r}$.

{\mb The limiting area fractions for $\ell(z)$ and $p(z)$ are treated separately. For $A_A=0$, no covered phase is present and we set $\ell(z)=p(z)=0$. For $A_A=1$, the lineal-path function is $\ell(z)=1$, while the chord-length density is set to $p(z)=0$. Only for $0<A_A<1$ are the closed-form expressions containing $L_A/A_A$ evaluated directly.}

{\mb
In addition to these established statistical descriptors, we use a derived scalar shape feature $\tau$. Its purpose is to distinguish grain shapes beyond aspect ratio. This is necessary because different grain families can have the same aspect ratio but still lead to systematically different effective material parameters.

The descriptor $\tau$ is defined as the normalized boundary-to-area ratio of the typical grain,
\begin{equation}
    \tau(\Xi_0)
    =
    \frac{\overline{L}(\Xi_0)/\overline{A}(\Xi_0)}
         {\overline{L}(\Xi_D)/\overline{A}(\Xi_D)} ,
\end{equation}
where $\Xi_D$ denotes the typical circular grain. This definition is motivated by the specific boundary length of the Boolean model. Since
\begin{equation}
    L_A =
    -\frac{\overline{L}}{\overline{A}}
    (1-A_A)\ln(1-A_A),
\end{equation}
the ratio $L_A(\Xi_0)/L_A(\Xi_D)$ is equal to $\tau(\Xi_0)$ for fixed area fraction. Thus, $\tau$ can be interpreted as the shape-dependent factor of the specific boundary length.

Although $\tau$ is closely related to $L_A$, it separates the shape-dependent information from the deterministic dependence of $L_A$ on the area fraction. This makes it a useful input feature when $A_A$ is already provided to the neural network.}

\section{Numerical Homogenization}

We consider two-phase hyperelastic composites. On the microscale, both phases are modeled by the isotropic Neo-Hooke energy density $\Psi$,
\begin{equation}
    \Psi = \frac{\mu}2(\mbox{tr}(\mathbf{F}^T\mathbf{F})-3)+\frac{\lambda}2\ln^2\mbox{det}(\mathbf{F}) - \mu\ln\mbox{det}(\mathbf{F}),
\end{equation}
where $\lambda$ and $\mu$ denote the Lam\'e parameters and $\mathbf{F}$ is the deformation gradient. The contrast $c$,
\begin{equation}
                c=\frac{\lambda_2}{\lambda_1}=\frac{\mu_2}{\mu_1}\in\{2,5,10,20,100\},
\end{equation}
is the ratio between the material parameters in the phases.

On the macroscale we represent the effective response by the same isotropic Neo-Hooke energy. Two limitations of this ansatz should be noted.
First, individual realizations of the random microstructure are generally not isotropic; average isotropic behavior can only be expected for ensembles of sufficiently many samples.
Second, the effective response cannot, in general, be represented by the model of the constituents. In the present setting, however, the fitted macroscopic model provides a sufficiently accurate approximation to the volume-averaged response, see \cite{2023Braendel}, p.~31.

For each microstructure realization in an ensemble, we solve a boundary value problem representing a uniaxial tensile test with $10\%$ prescribed deformation under Dirichlet boundary conditions. The nonlinear problem is solved by an adaptive load-stepping scheme, and the resulting first Piola-Kirchhoff stress and deformation gradient are volume-averaged at ten standardized load levels.

The effective Lamé parameters of the macroscopic Neo-Hookean model are fitted to the ensemble-averaged, volume-averaged response using the Equilibrium Gap Method. In this least-squares approach, the error between the measured macroscopic stress response and the stress computed from the assumed Neo-Hookean model for the corresponding measured deformation gradients is minimized.

\subsection{Size of the RVE (Representative Volume Element)}

To interpret the computed macroscopic response as an effective material response, the chosen volume element has to be representative of the underlying random microstructure. This involves two aspects: the volume element has to be statistically typical of the material, and it has to be sufficiently large such that the apparent response is only weakly affected by the imposed boundary conditions.

For the stationary Boolean model considered here, statistical typicality is addressed by sampling independent realizations of an ergodic random medium. For finite volume elements, however, statistical fluctuations remain. We therefore assess the required volume-element size following the statistical approach of \cite{2003Kanit}. Ensembles of increasing volume-element size $\delta$ are considered, and the RVE size $\delta_{\mathrm{RVE}}$ is chosen such that the prescribed accuracy of the ensemble-averaged macroscopic response is reached.

In the present study, the criterion is applied to the volume-averaged first Piola-Kirchhoff stress $\mathbf{P}$. For an ensemble size $n=200$, a prescribed relative error $\varepsilon(\langle \mathbf{P}_{11}\rangle)=0.01$, and a confidence level of $95\%$, only ensembles with volume elements with $\delta > \delta_{\mathrm{RVE}}$ are used for the computation of the effective material response. Note that that $\delta_{\mathrm{RVE}}$ differs strongly between different grain types and different contrasts, see \cite{2023Braendel}, pp.~124.

\section{Machine Learning Surrogate Model}

To predict the effective Lam\'e parameters from microstructural information, we employ a fully connected multilayer perceptron (MLP) with three hidden layers of 512 neurons each. We use ReLU activation functions. The network maps a feature vector assembled from scalar statistical descriptors, material properties, and sampled curve-valued statistical descriptors to two outputs corresponding to the effective Lam\'e parameters. {\mb We use PyTorch \cite{2024PyTorch} version 2.9.0+cu130.}

Before training, the input features are standardized by subtracting the mean and dividing by the standard deviation, while the target values {\mb $\lambda,\mu$} are {\mb first} contrast-normalized {\mb $\lambda_n,\mu_n \in[0,1]$} using min-max scaling with respect to the corresponding phase parameters {\mb and then standardized $z_{\lambda_n},z_{\mu_n}$ as well}.
This preprocessing improves the numerical stability of the optimization. The target parameters are transformed back to their physical range only for evaluation of the prediction error.

The distance-dependent descriptors $S_2(r)$, $\ell(z)$, and $p(z)$ are not passed to the network as full curves. Instead, they are sampled at a prescribed number of quadratically distributed points on their respective grids. More precisely, {\orh for a grid variable $t=r$ or $t=z$, and
$t \in[t_{\min},t_{\max}]$,} the sampling positions are chosen as
\begin{equation}
    t_i = t_{\min} + (t_{\max}-t_{\min})\left(\frac{i}{m-1}\right)^2,
    \qquad i=0,\dots,m-1,
\end{equation}
where $m$ denotes the number of curve samples. In the computations reported here, we use $m=16$.

We {\mb use a leave-one-grain-type-out cross-validation strategy (LOGO-CV). In each LOGO-CV} fold, all samples corresponding to one grain type are excluded from training and are used as unseen-type test data. The remaining samples are randomly split into 85\% training data and 15\% validation data. {\mb The model trained in such a fold is refered to as a {\it fold predictor}.} This setup assesses the ability of the surrogate to generalize to microstructures with previously unseen grain geometries.
{\mb For a fixed feature set, we additionally define an {\it ensemble predictor} as the unweighted average of the predictions of all fold predictors. This ensemble predictor is used as an aggregated diagnostic predictor and to visualize smooth response surfaces. Its error should therefore not be interpreted as a strict unseen-type error: for a given grain type, the corresponding samples are unseen only for the fold predictor in which that grain type was held out, while they are present in the training data of the other fold predictors. Strict unseen-type performance is reported separately by the individual \textit{grain-type-}fold errors.}

The networks are trained for {\mb $1000$} epochs using a batch size of {\mb $N=128$}.
Optimization is performed with Adam and a learning rate $\alpha=1\cdot10^{-5}$.

{\mb Let $q\in\{\lambda_n,\mu_n\}$ denote a contrast-normalized target and let $z_q\in\{z_{\lambda_n},z_{\mu_n}\}$ be the corresponding standardized quantity. The network predicts $\hat z_q$. The corresponding prediction transformed back to the contrast-normalized target space is denoted by $\hat q$.
The loss $\mathcal{L}$ consists of a mean squared error in the standardized target space and an additional soft-relative penalty in the contrast-normalized target space,
\begin{equation}
    \mathcal{L} = \frac{1}{2N} \sum_{i=1}^{N} \sum_{q\in\{\lambda_n,\mu_n\}} \left[ \left(\hat{z}_{q,i}-z_{q,i}\right)^2
    + \beta \left( \frac{ \hat{q}_i-q_i }{ \max\left(|q_i|+\gamma,\varepsilon\right) } \right)^2 \right].
\end{equation}
Here, $\beta$ controls the strength of the soft-relative contribution, $\gamma>0$ reduces the sensitivity of the relative penalty near zero, and $\varepsilon>0$ is used for numerical stability. In this work we choose $\beta=0.15$, $\gamma=0.01$, and $\varepsilon=1\cdot10^{-8}$.}

\subsection{Datasets}
The primary dataset consists of 506 samples of effective Lam\'e parameters obtained from numerical homogenization of ensembles of two-phase planar Boolean model microstructures \cite{2023Braendel}. These samples are given for 9 different types of grains, 5 different contrasts $c\in\{2,\ 5,\ 10,\ 20,\ 100\}$ in the microscopic phase material parameters, and 11 different fractions $A_A\in\{0,\ 0.1,\ \ldots,\ 1\}$ of the Boolean model.

Out of the total of 506 samples, 11 samples represent separate computations involving longer rectangles with an aspect ratio of $1:16$, with the same set of fractions but only a contrast of $c=100$.

A second dataset is used as a synthetic boundary augmentation and consists of 9010 samples. It contains only the limiting fractions $A_A\in\{0,\ 1\}$ for the same 5 different contrasts and a set of the scalar descriptor $\tau\in\{1,\ 1.01,\ \ldots,\ 10\}$.

{\mb For visualization, additional inference grids are used after training. The normalized prediction curves in Figure~\ref{fig:normalized_prediction} are evaluated on a dense grid of area fractions $A_A\in\{0,\ 0.01,\ldots,\ 1\}$ for the sampled contrast values, while the heatmaps in Figure~\ref{fig:heatmap-d} are evaluated on a dense grid in the $(A_A,c)$-plane. These points are not used for training or for the reported error statistics, but only to visualize the learned response surfaces and to inspect the behavior between sampled parameter values.}

\section{Results}

{\mb Table~\ref{tab:rel_errors} reports summary statistics of the empirical relative-error distributions.
The absolute relative errors are computed in physical parameter space and pooled over the two target quantities $\lambda$ and $\mu$. Here, $P_q$ denotes the $q$th percentile of the empirical relative-error distribution.
We use the $95$th percentile $P_{95}$ as the primary ranking criterion, since it characterizes the upper tail of the error distribution while being less sensitive to single outliers than the maximum.
For each feature set, the ensemble predictor error and the two largest grain-type-fold errors are reported.
Across the investigated feature sets, the $D$-fold gives the largest upper-tail unseen-type error. The smallest worst-fold $P_{95}$ value is obtained for the feature set $(A_A,c,\tau,S_2(r),\ell(z))$.
However, dense post-training response-surface evaluations revealed non-admissible contrast-normalized predictions for $A_A\in[0.95,0.99]$, where $\hat\lambda_n,\hat\mu_n>1$. We therefore use the slightly simpler feature set $(A_A,c,\tau,S_2(r))$ for the remainder of this section.}

\begin{figure}[htb]
\begin{center}
\begin{minipage}[t]{\linewidth}
\centering
\includegraphics[height=80mm]{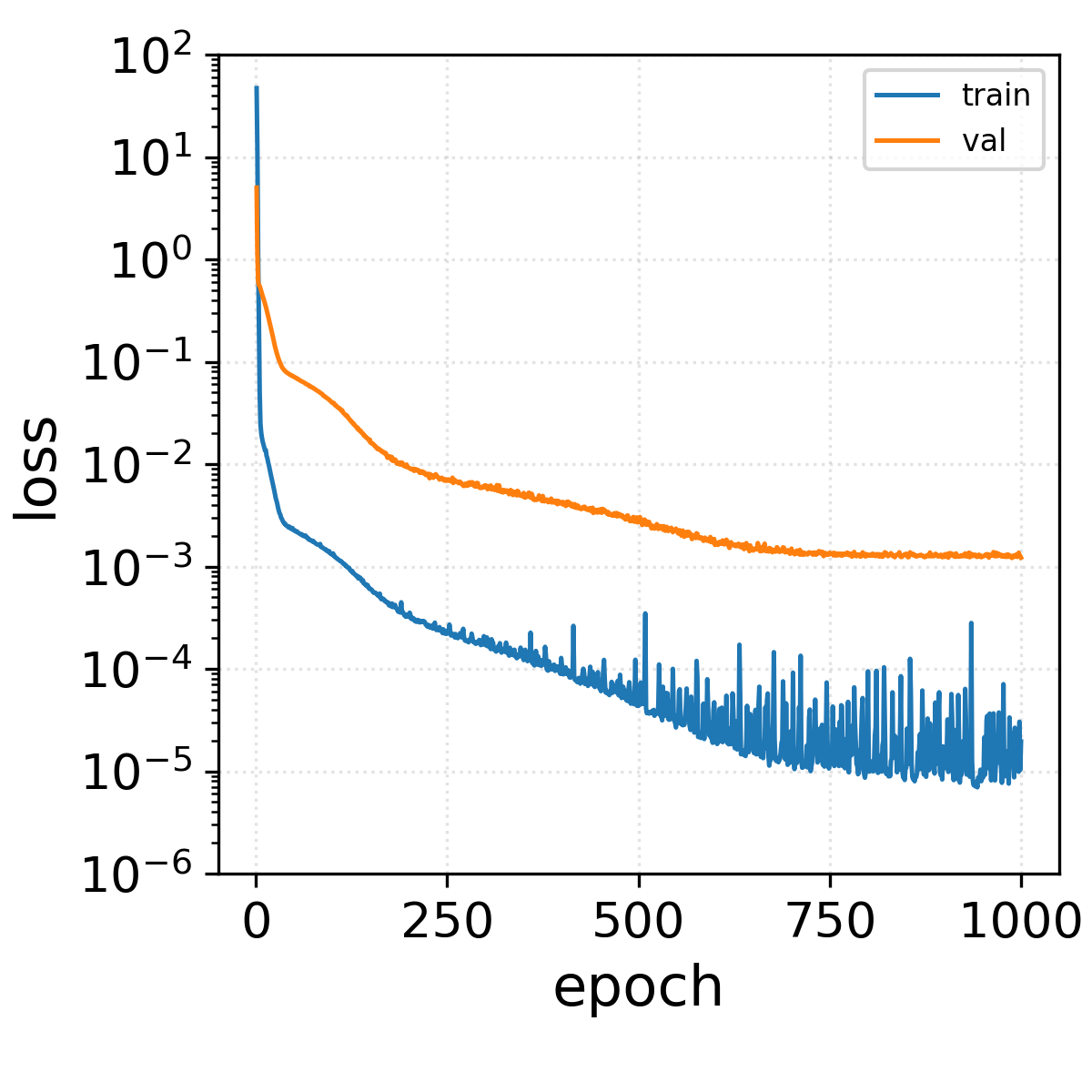}
\caption{\mb Training and validation loss history for the feature set $(A_A,c,\tau,S_2(r))$ in the $D$-fold.}
\label{fig:history}
\end{minipage}
\end{center}
\end{figure}

{\mb The loss history in Figure~\ref{fig:history} is typical for the recorded runs and shows a stable decrease of both training and validation loss.}

\begin{figure}[htb]
\begin{center}
\begin{minipage}[t]{\linewidth}
\centering
\includegraphics[height=75mm]{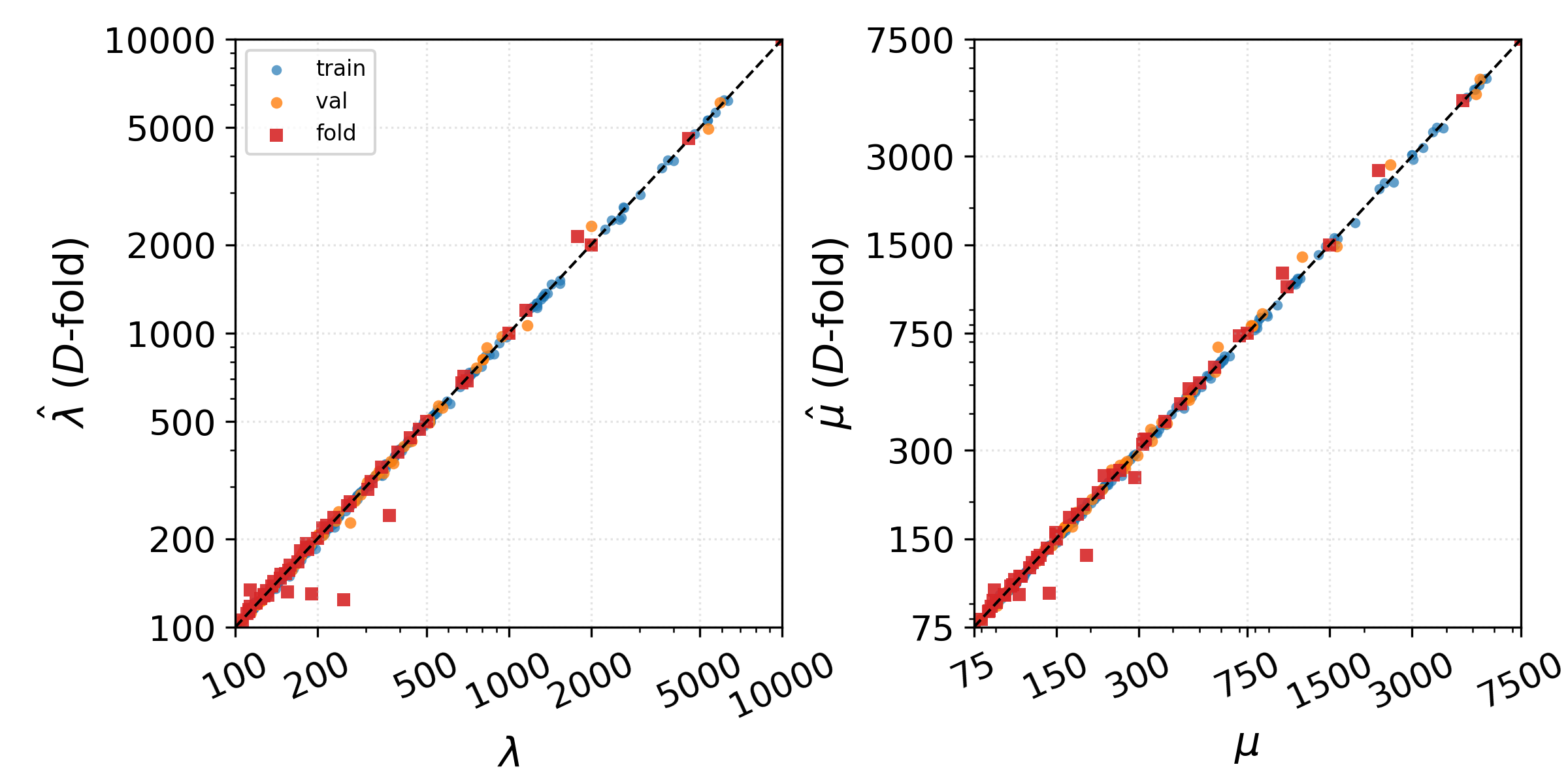}\\
\includegraphics[height=75mm]{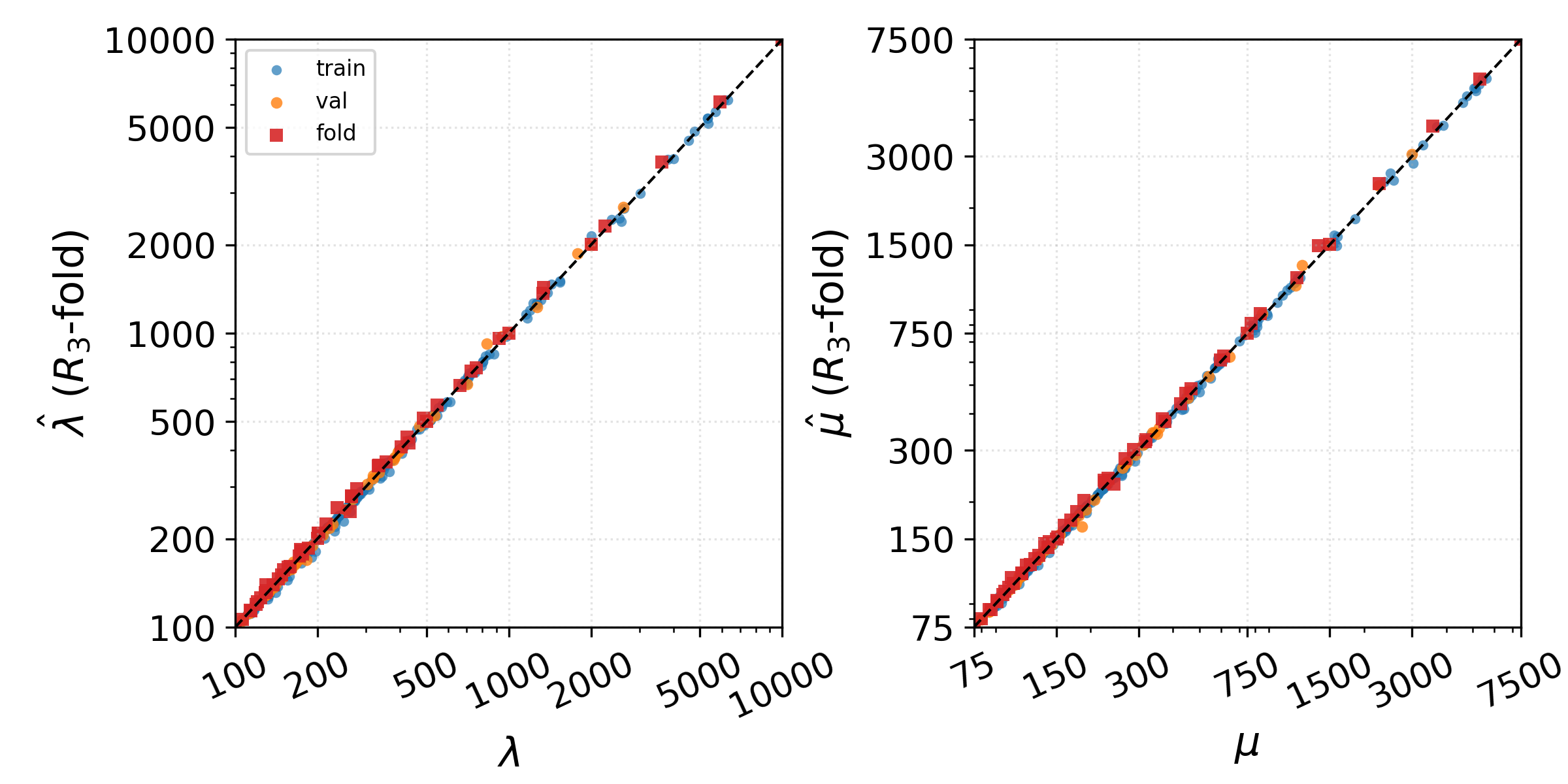}
    \caption{\mb Parity plots for the predicted and reference Lamé parameters $\lambda$ and $\mu$ for the feature set $(A_A,c,\tau,S_2(r))$. {\orh The upper pair of plots shows the challenging $D$-fold ($P_{95}=0.195$), while the lower pair shows the better-performing $R_3$-fold ($P_{95}=0.066$).} Blue and orange markers denote training and validation samples, respectively. {\orh Red markers denote the held-out grain type of the corresponding fold, i.~e., $D$ on the top and $R_3$ on the bottom.} The dashed line indicates perfect prediction.}
\label{fig:inference-comparison}
\end{minipage}
\end{center}
\end{figure}

{\orh Figure~\ref{fig:inference-comparison} compares parity plots for two LOGO-CV folds of the selected feature set $(A_A,c,\tau,S_2(r))$: the $D$-fold, which has the largest unseen-type error for this feature set, and the $R_3$-fold, which serves as a substantially better-performing comparison case. In both folds, the training and validation samples are predicted close to the diagonal, while the held-out samples reveal the difference in unseen-type generalization quality.}

\begin{figure}
\begin{center}
\begin{minipage}{\linewidth}
    \includegraphics[width=\linewidth]{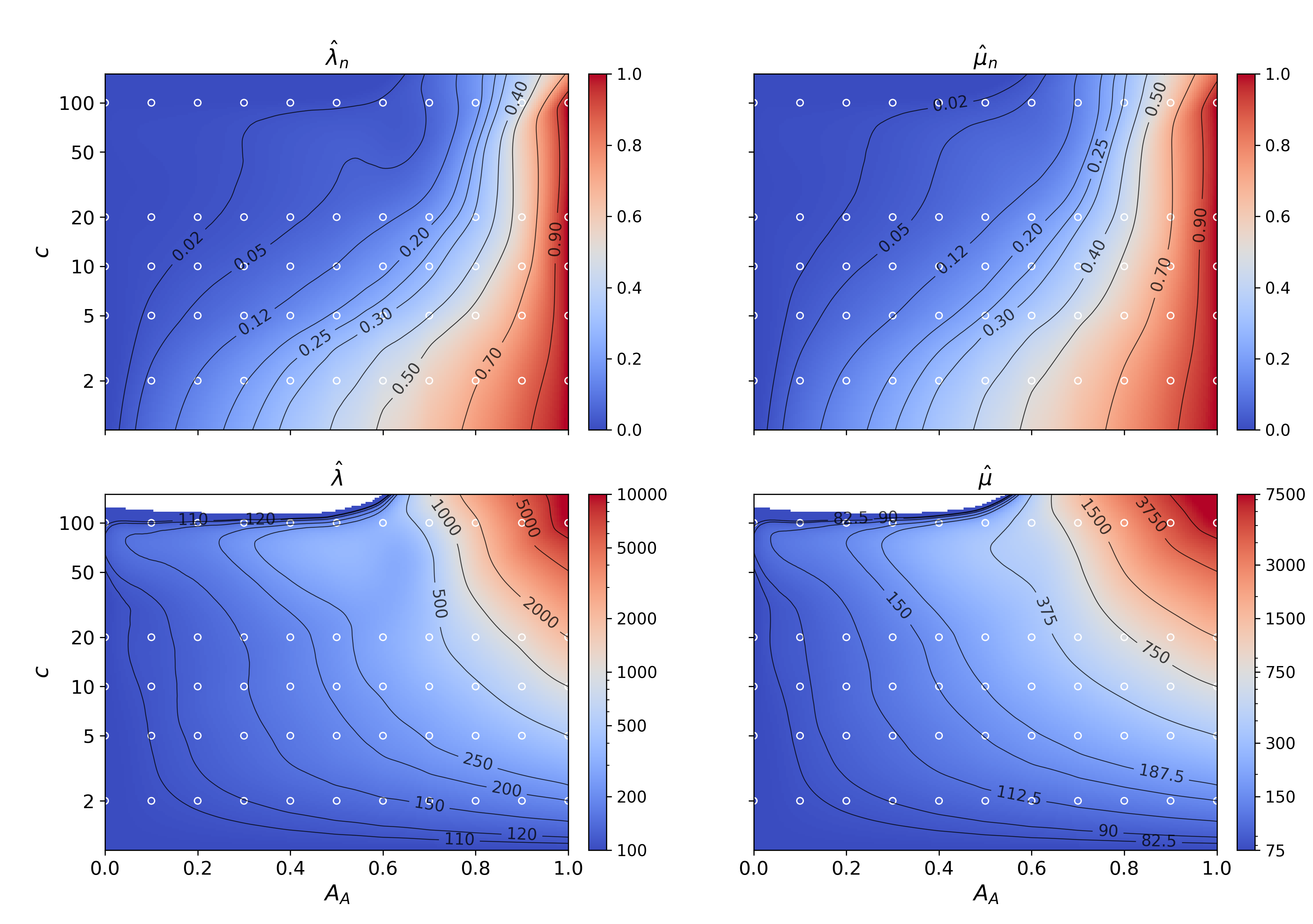}
    \caption{\mb Predicted response surfaces of the ensemble predictor for the feature set $(A_A,c,\tau,S_2(r))$ for the grain type $D$. The upper row shows the contrast-normalized predictions $\hat \lambda_n$ and $\hat \mu_n$, while the lower row shows the corresponding predicted Lamé parameters $\hat\lambda$ and $\hat\mu$. The predictions are evaluated on a dense grid in area fraction $A_A$ and contrast $c$. White markers indicate the sampled parameter combinations used in the data set.}
\label{fig:heatmap-d}
\end{minipage}
\end{center}
\end{figure}

{\mb Figure~\ref{fig:heatmap-d} visualizes the response surface predicted by the ensemble model. Within the sampled part of the parameter domain, the ensemble prediction varies smoothly with both area fraction and contrast. In contrast, deviations become visible in sparsely sampled and extrapolated regions, in particular between the largest sampled contrast levels and beyond the training range. This is most visible for the denormalized Lamé parameters, where small deviations in the contrast-normalized targets are amplified by the contrast-dependent denormalization.}

\begin{figure}
\begin{center}
\begin{minipage}{\linewidth}
    \includegraphics[width=\linewidth]{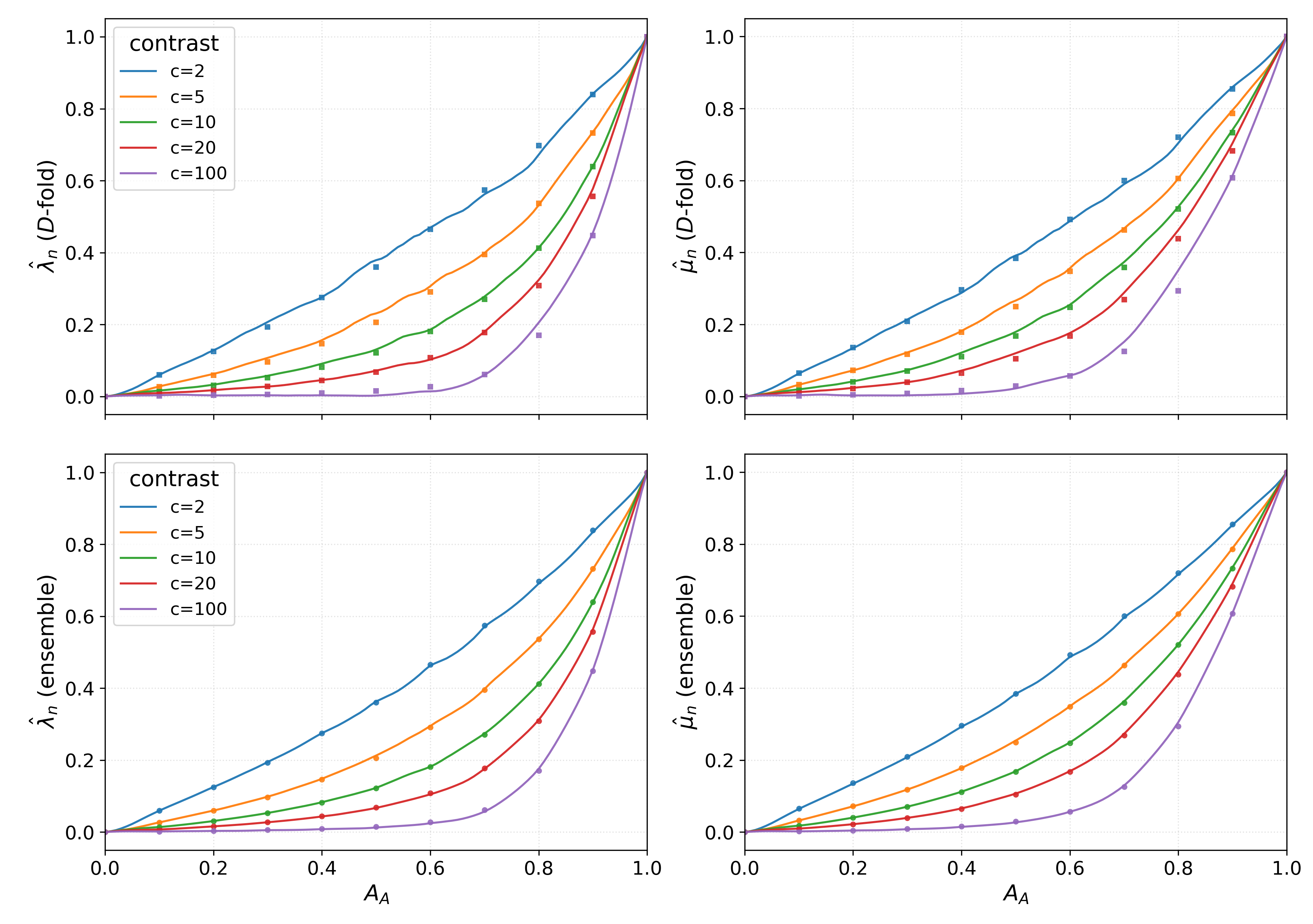}
\caption{\mb Predicted contrast-normalized Lamé parameters for grain type $D$ using the selected feature set $(A_A,c,\tau,S_2(r))$. The upper row shows the prediction of the $D$-fold model, while the lower row shows the corresponding ensemble prediction. The columns show $\hat{\lambda}_n$ and $\hat{\mu}_n$ as functions of the area fraction $A_A$ for different contrast values. Marker shapes refer to the predictor shown in the corresponding row: squares denote held-out unseen-type samples for the $D$-fold predictor, whereas circles denote samples that are available to the other fold predictors contributing to the ensemble predictor.}
\label{fig:normalized_prediction}
\end{minipage}
\end{center}
\end{figure}

{\mb Figure~\ref{fig:normalized_prediction} shows the predicted contrast-normalized Lamé parameters for grain type $D$ as functions of the area fraction. The upper row corresponds to the $D$-fold predictor and the lower row shows the ensemble predictor.
The ensemble prediction provides smoother response curves, whereas the individual $D$-fold predictor exhibits stronger local variations. Still, the fold predictor reproduces the expected qualitative dependence on area fraction and contrast and remains within the admissible normalized range.}
{\orh The observed small-scale waviness in Figure~\ref{fig:normalized_prediction} indicates that the purely data-driven model does not explicitly enforce monotonicity or smoothness with respect to the area fraction. While this has only a minor effect on the quantitative accuracy at the sampled data points, it suggests that additional structure-preserving constraints could improve the physical consistency of the predictions.}

\begin{table}
\begin{center}
\begin{minipage}{\linewidth}
\centering
\begin{tabular}{@{}lrcccc@{}}
\hline
    \textbf{Feature set} & \textbf{Evaluation} & \textbf{Mean} & \textbf{Median} & \textbf{$P_{90}$} & \textbf{$P_{95}$} \\
\hline
\hline
     & ensemble   & 0.009 & 0.003 & 0.021 & 0.036 \\
    $(A_A, c, \tau, L_A, \chi_A, S_2(r), \ell(z), p(z))$ & $D$-fold   & 0.216 & 0.027 & 0.365 & \textit{1.596} \\
     & $Q$-fold   & 0.040 & 0.009 & 0.134 & 0.196 \\
\hline
                                & ensemble   & 0.012 & 0.004 & 0.029 & 0.044 \\
    $(A_A, c, \tau, p(z))$                            & $D$-fold   & 0.251 & 0.048 & 0.833 & \textit{1.434} \\
                                & $Q$-fold   & 0.055 & 0.017 & 0.180 & 0.261 \\
\hline
                   & ensemble   & 0.021 & 0.005 & 0.058 & 0.096 \\
    $(A_A, c, L_A, S_2(r), \ell(z), p(z))$               & $D$-fold   & 0.126 & 0.034 & 0.422 & \textit{0.642} \\
                   & $Q$-fold   & 0.071 & 0.031 & 0.169 & 0.369 \\
\hline
                                            & ensemble   & 0.057 & 0.015 & 0.153 & 0.255 \\
    $(A_A, c)$                                        & $D$-fold   & 0.134 & 0.051 & 0.301 & \textit{0.632} \\
                                            & $E_3$-fold & 0.106 & 0.049 & 0.295 & 0.477 \\
\hline
                                       & ensemble   & 0.036 & 0.012 & 0.096 & 0.151 \\
    $(A_A, c, L_A)$                                   & $D$-fold   & 0.110 & 0.062 & 0.272 & \textit{0.475} \\
                                       & $Q$-fold   & 0.080 & 0.024 & 0.222 & 0.436 \\
\hline
                                    & ensemble   & 0.018 & 0.006 & 0.047 & 0.069 \\
    $(A_A, c, \chi_A)$                                & $D$-fold   & 0.098 & 0.034 & 0.223 & \textit{0.413} \\
                                    & $S$-fold   & 0.055 & 0.015 & 0.160 & 0.281 \\
\hline
                         & ensemble   & 0.009 & 0.004 & 0.025 & 0.039 \\
    $(A_A, c, \tau, L_A, \chi_A)$                     & $D$-fold   & 0.087 & 0.067 & 0.172 & \textit{0.306} \\
                         & $R_3$-fold & 0.056 & 0.038 & 0.148 & 0.174 \\
\hline
                                      & ensemble   & 0.009 & 0.004 & 0.024 & 0.036 \\
    $(A_A, c, \tau)$                                  & $D$-fold   & 0.048 & 0.016 & 0.108 & \textit{0.282} \\
                                      & $S$-fold   & 0.028 & 0.011 & 0.070 & 0.133 \\
\hline
                                & ensemble   & 0.007 & 0.003 & 0.021 & 0.031 \\
    $(A_A, c, \tau, \ell(z))$                            & $D$-fold   & 0.040 & 0.009 & 0.158 & \textit{0.255} \\
                                & $S$-fold   & 0.032 & 0.013 & 0.101 & 0.144 \\
\hline
                              & ensemble   & 0.008 & 0.003 & 0.022 & 0.032 \\
    $(A_A, c, \tau, S_2(r))$                          & $D$-fold   & 0.042 & 0.012 & 0.126 & \textit{0.195} \\
                              & $S$-fold   & 0.030 & 0.006 & 0.084 & 0.142 \\
\hline
                       & ensemble   & 0.006 & 0.002 & 0.019 & 0.026 \\
    $(A_A, c, \tau, S_2(r), \ell(z))$                   & $D$-fold   & 0.034 & 0.016 & 0.086 & \textit{0.171} \\
                       & $S$-fold   & 0.029 & 0.013 & 0.076 & 0.112 \\
\hline
\hline
\end{tabular}
\caption{\mb Relative error statistics for the LOGO-CV experiments.
    {\orh The entries \textit{grain-type}-fold refer to predictors trained without the corresponding grain type samples. Their errors quantify generalization to the unseen grain type. The entries ensemble refer to the aggregated prediction of all fold predictors of the feature set.}
 $P_q$ denotes the $q$th percentile of the empirical relative-error distribution. Feature sets are sorted by the largest unseen-type $P_{95}$ error in descending order; the corresponding values are in italics.
For each feature set, the ensemble error and the two largest unseen-type fold errors are reported.
}
\label{tab:rel_errors}
\end{minipage}
\end{center}
\end{table}

{\mb On an NVIDIA RTX~5090, training a single fold predictor takes approximately one minute. The subsequent inference time is negligible compared with both the training time and the numerical homogenization required to generate the data.}

\paragraph{Use of artificial intelligence tools.}
OpenAI ChatGPT-5.4 was used in May 2026 as an editorial and coding-assistance tool during the preparation of this manuscript. Its use was limited to language polishing, consistency checks, and assistance in generating preliminary Python evaluation scripts. It was not used to generate the numerical homogenization data, train or evaluate the surrogate models, or determine the scientific claims and conclusions. All AI-assisted suggestions were reviewed, edited, and verified by the authors, who take full responsibility for the final manuscript.

\section{Conclusion}
{\mb
Descriptor-based neural-network surrogates provide an efficient approximation of effective Lamé parameters for hyperelastic Boolean-model microstructures.
The LOGO-CV results show that statistical descriptors improve the prediction of unseen grain geometries. In particular, scalar shape information together with the two-point correlation function gives a compact feature representation with good sampled error statistics and regular response behavior.

The results also indicate that the lineal-path function is a promising additional descriptor. The feature set $(A_A,c,\tau,S_2(r),\ell(z))$ gives the smallest sampled worst-fold error among the investigated descriptor combinations.
However, dense post-training response evaluations reveal that this improvement in sampled error does not automatically imply a physically admissible response surface between sampled parameter values.

The LOGO-CV setting further shows that unseen-type generalization remains the central difficulty of the surrogate model.
The $D$-fold consistently appears as the most challenging split among the descriptor combinations investigated in this work. One possible reason is that the corresponding effective parameters, and especially their contrast-normalized values, lie close to the lower end of the admissible range, where small absolute deviations can lead to large relative errors.

Future work should therefore investigate loss formulations and bounded output parametrizations that enforce admissibility of the contrast-normalized targets.
In addition, future models should include constraints or regularization terms that promote smooth dependence on area fraction and contrast and consistent endpoint behavior. Together with denser sampling of difficult high-contrast regions and a systematic study of the sampling and representation of curve-valued descriptors, this may improve the robustness and physical consistency of unseen-type predictions.
}

\section*{Acknowledgements}
The authors acknowledge computing time on the compute cluster of the Faculty of Mathematics and Computer Science of Technische Universität Bergakademie Freiberg, operated by the computing center (URZ) and funded by the Deutsche Forschungsgemeinschaft (DFG) under DFG grant number 397252409 (\url{https://gepris.dfg.de/gepris/projekt/397252409}).

\vspace{\baselineskip}

\end{document}